\documentclass[sw]{AGUTeX}

\usepackage{graphicx}

\authorrunninghead{MARSH ET AL.}

\titlerunninghead{SPARX}

\begin{document}

%
%

\title{SPARX: a modelling system for Solar Energetic Particle Radiation Space Weather forecasting}
%
%

%
%



\authors{M. S. Marsh,\altaffilmark{1,2} S. Dalla,\altaffilmark{1} M. Dierckxsens,\altaffilmark{3} T. Laitinen,\altaffilmark{1} and N. B. Crosby,\altaffilmark{3}}

\altaffiltext{1}{Jeremiah Horrocks Institute,
University of Central Lancashire, Preston, UK.}

\altaffiltext{2}{Met Office, Exeter, Devon, UK.}

\altaffiltext{3}{Belgian Institute for Space Aeronomy (BIRA-IASB),
Brussels, Belgium.}

\begin{abstract}
The capability to predict the parameters of an SEP event such as its onset, peak flux, and duration is critical to assessing any potential space weather impact.
We present a new flexible modelling system simulating the propagation of Solar Energetic Particles (SEPs)
from locations near the Sun to any given location in the heliosphere to forecast the SEP flux profiles. 
SPARX uses an innovative methodology, that allows implementation within an operational framework, to overcome the time constraints of test particle modelling of SEP profiles, allowing the production of near real time SEP nowcasts and forecasts, when paired with appropriate near real time triggers. SPARX has the capability to produce SEP forecasts within minutes of being triggered by observations of a solar eruptive event.
The model is based on the test particle approach and
is spatially 3D, thus allowing for the possibility of transport in the direction 
perpendicular to the magnetic field. The model naturally includes the effects of perpendicular propagation due to drifts and drift-induced deceleration.
The modelling framework and the way in which parameters of relevance
for Space Weather forecasting are obtained are described.
The first results from the modelling system are presented. These results demonstrate that corotation 
and drift of SEP streams play an important role in shaping SEP flux profiles. 

\end{abstract}

%
%

%

\begin{article}

%
%

\section{Introduction}\label{sect.intro}

In recent years, much work within Space Weather (SWx) has focussed on the problem
of forecasting Solar Energetic Particle (SEP) intensities near Earth
and at other locations in the heliosphere, following the observation of a
solar eruptive event, such as a solar flare or Coronal Mass Ejection (CME).

The flux of ionising radiation due to SEPs may increase significantly within regions of space, following a solar event. The resulting absorbed dose has a range of biological and technological impacts, the most severe of which is the increased radiation dose faced by astronauts in low Earth orbit or on future interplanetary missions \citep{Nap2008}. Severe SEP events may also increase the radiation dose received at aviation altitudes and must be considered in the assessment of risk posed to passengers and crew \citep{Ams2007}; additionally, the use of transpolar flight paths may be affected due to the disruption of high frequency communications used by airlines. SEP events may also impact the operation of satellites causing hardware upsets in sensitive electronics, disruption of optical star trackers used for spacecraft orientation \citep{Bak2002}, and degradation of solar panel performance \citep{Bre2005}. 
It is therefore important to know when a SEP radiation event will begin, when dangerous levels will be exceeded and how long it will persist.

Several operational empirical tools for predicting SEP events exist, including 
the SWPC SEP prediction model \citep{Bal2008}, RELEASE \cite{pos2007}, MAG4 \citep{Fal2011}, UMASEP \cite{Nun2011} and the Proton Prediction System \citep{Kah2007},
and other similar schemes have been proposed \citep{Lau2009}.
In recent years, considerable effort has been devoted to introducing physical
models of SEP propagation within forecasting systems.
A number of challenges face SEP modellers, due to the lack of direct information on the
energetic particle properties, location of acceleration (due to a flare or 
CME-driven shock), the complexity of the physics of their
propagation in the 3D turbulent heliospheric plasma, and the computational expense considering the timescales required to produce an actionable forecast.

Sophisticated Space Weather models for SEP forecasting have been developed recently.
SOLPENCO uses MHD simulations of the propagation of a CME-driven shock to model accelerated SEPs, by solving the focussed transport equation  \citep{Ara2006}.
\cite{Luh2010} predict the interplanetary accelerated component of SEP fluxes by using a cone model description of the CME shock and propagating the injected particles scatter-free, making use of conservation of angular momentum.
The EPREM module within the EMMREM framework solves a diffusion equation including adiabatic deceleration and the effect of
pitch angle averaged drifts \citep{Sch2010, Koz2010}.
All of these tools are broadly based on the solution of a transport equation for the distribution
function of the energetic particles. The majority assume that one spatial variable, the distance travelled along the magnetic field line, is sufficient for a description of the problem.
In other words, particle propagation is assumed to take place along the magnetic field
only, with no significant perpendicular transport.

This paper presents a new fully-3D physics based model for simulating SEP propagation for Space Weather forecasting purposes: Solar Particle Radiation SWx (SPARX).
The model is the first one to use the test particle approach within a forecasting context and naturally describes perpendicular transport effects due to particle drift. Unlike many other Space Weather SEP modelling frameworks, it makes use of a full-orbit test particle 
approach to integrate trajectories of a large number of particles and produce time
dependent particle flux profiles at a given location.
SPARX was initially developed as part of the European Union-funded FP7 COMESEP project \citep{Cro2012} and is operational within its SEP forecasting component which produces alerts at www.comesep.eu/alert .

In this paper we describe the methodology behind the model and forecasting system, an example of how the test particle approach has been implemented in an operational context, within the COMESEP Alert System, and present examples of its output.
The output of the model displays a number of features that are unique to the approach used, and that are not present in standard focussed transport modelling.

\section{Model} \label{sec.model}
The numerical model is based upon a relativistic full-orbit test particle code that was originally developed to study 
particle acceleration during magnetic reconnection \citep{Dal2005} and was modified to describe the
heliospheric propagation problem \citep{Mar2013}. 
Each simulation follows a large number $N$ of independent test particles. For each of them,
the equation of motion:
\begin{equation}  
\frac{d {\bf p}}{d t}  = q \, \left( {\bf E} + \frac{1}{c}  \frac{{\bf p}}{m_0 \gamma} \times {\bf B} \right),\label{eqn_motion}
\end{equation}
is solved numerically, where ${\bf p}$ is the particle's momentum, $t$ is time, $q$ the particle charge, $m_0$ its rest mass, $\gamma$ its Lorentz factor and $c$ is the speed of light.
The electric field ${\bf E}$ and magnetic field ${\bf B}$ in which the particles propagate are defined as those of a unipolar Parker spiral:
\begin{eqnarray}
\mathbf{B} & = & P\left(\theta\right) \, \left(\frac{B_{0} r_{0}^{2}}{r^{2}} \, \mathbf{e}_{r} - \frac{B_0  \, r_0^2 \, \Omega}{v_{sw}} \, \frac{\sin{\theta}}{r} \, \mathbf{e_{\phi}} \right), \label{eqn_B}\\
\mathbf{E} & = & P\left(\theta\right) \, \left( - \frac{\Omega  \, B_0 r_0^2}{c}   \,  \frac{\sin{\theta}}{r} \, \mathbf{e_{\theta}} \right), \label{eqn_E}
\end{eqnarray} 
where $(r, \theta, \phi)$ are spherical coordinates giving radial distance, colatitude and longitude respectively, $B_{0}$ is the radial magnetic field magnitude at a reference surface of radius $r_{0}$ defined so as to give a magnetic field strength of $5$~nT at $1$~AU,  $\Omega$ is the solar angular rotation rate, $v_{sw}$ is the solar wind speed and the $\mathbf{e}$ parameters represent standard unit vectors. 
The values of the constants are as described in \cite{Mar2013}. 
The function $P\left(\theta\right)$ in Eqs.(\ref{eqn_B})--(\ref{eqn_E}) determines the sign of the magnetic and electric field used to produce forecasts; this is dependent upon the event colatitude and magnetic phase of the solar cycle, as described in Section~\ref{sec.tiling}. 

Eqs.(\ref{eqn_motion})--(\ref{eqn_E}) describe the particle motion in a fixed reference frame that sees the solar wind move radially outwards with
speed  $v_{sw}$ and the Sun rotate with angular rate  $\Omega$. In this frame, the effects of solar rotation are included and the SEP parameters derived can be directly compared with those measured by spacecraft.
The effect of turbulence in the heliosphere is modelled by introducing isotropic scattering of each particle's pitch angle and gyro-phase in the reference frame of the solar wind, using a Poisson distributed series of scattering times, determined by a prescribed value of mean free path
$\lambda$ \citep{Mar2013}.

In its present form, SPARX describes the SEP injection taking place instantaneously near the Sun, from a source region that is spatially extended, representing a CME driven shock in the corona. It does not
model any additional spatially and time dependent injection that can take place at CME-driven shocks in interplanetary space.
This is likely to be a reasonable approximation for large SEP energies (eg $>$60 MeV), for which most
of the acceleration is thought to take place near the Sun. An extension of the model to include a long duration injection will be carried
out in the future.

Particles are injected into the simulation at  $r=2$ solar radii, with a
power law energy spectrum $E$$^{-\gamma}$ over the energy range 10--400 MeV, where the spectral index $\gamma$ may be selected.
Their initial velocity vectors are randomly distributed in a semi-hemisphere and oriented outward from the Sun.
Concerning the spatial distribution of the injected particles, to enhance computational efficiency, an extended injection region is constructed as a composite of smaller injection region `tiles'.
The size of the extended injection region can vary between a minimum size of 6$^{\circ}$$\times$6$^{\circ}$ and a larger size 
that is derived by combining multiple 6$^{\circ}$$\times$6$^{\circ}$ injection \lq tiles\rq, with the aim of representing
a CME-driven shock region at the Sun.
During the simulation, each time a particle crosses a sphere of radius 1 AU its parameters at this time 
are output such as time, longitude, latitude, kinetic energy and pitch angle. This information is used to produce synthetic particle flux profiles to forecast what would be observed by a spacecraft at 1~AU. 

Results from the full orbit test particle model have shown that large scale drifts
associated with the Parker spiral magnetic field play an important role in SEP propagation \citep{Mar2013, Dal2013}. 
Drifts cause perpendicular propagation, by moving particles away from the magnetic field line on which they were 
originally injected. This effect, as measured in terms of displacement from the injection
field line, is most prominent for partially ionised heavy ions and for protons at
the high end of the SEP energy range \citep{Mar2013}.
In addition, drift produces very strong particle deceleration (Dalla et al., Drift induced
deceleration of Solar Energetic Particles, submitted to {\it Astrophysical Journal}, 2015), so that the energy of an SEP
near-Earth is often much smaller than the energy with which it left the Sun. This effect, which is naturally
described within SPARX, influences the predicted particle flux profiles within a given energy range.

The fact that drifts are significant for SEPs
means that the polarity of the interplanetary magnetic field (IMF) along a particle's 
trajectory plays an important role, as it determines whether the drift in the latitudinal direction will
be positive or negative. The heliospheric current sheet also introduces additional particle drift.
Therefore in principle the IMF polarity, the current sheet location, and previous disturbances to the IMF geometry are required in 3D for accurate modelling. 
In the current version of the model, for operational running, we do not attempt to use as input a complete description of
these quantities. The large scale structure of the IMF is approximated
using a unipolar model with the polarity determined by the hemisphere in which the solar eruptive event occurs, and the prevalent A$^+$ or A$^-$ solar cycle phase at the time of the event, where the A$^+$ or A$^-$ patterns are those defined in the analysis of galactic cosmic ray drifts, described in Section~\ref{sec.tiling}.

\section{Operational running}  \label{sec.oper_running}

\begin{figure}[h]
\centering
\includegraphics[width=19pc]{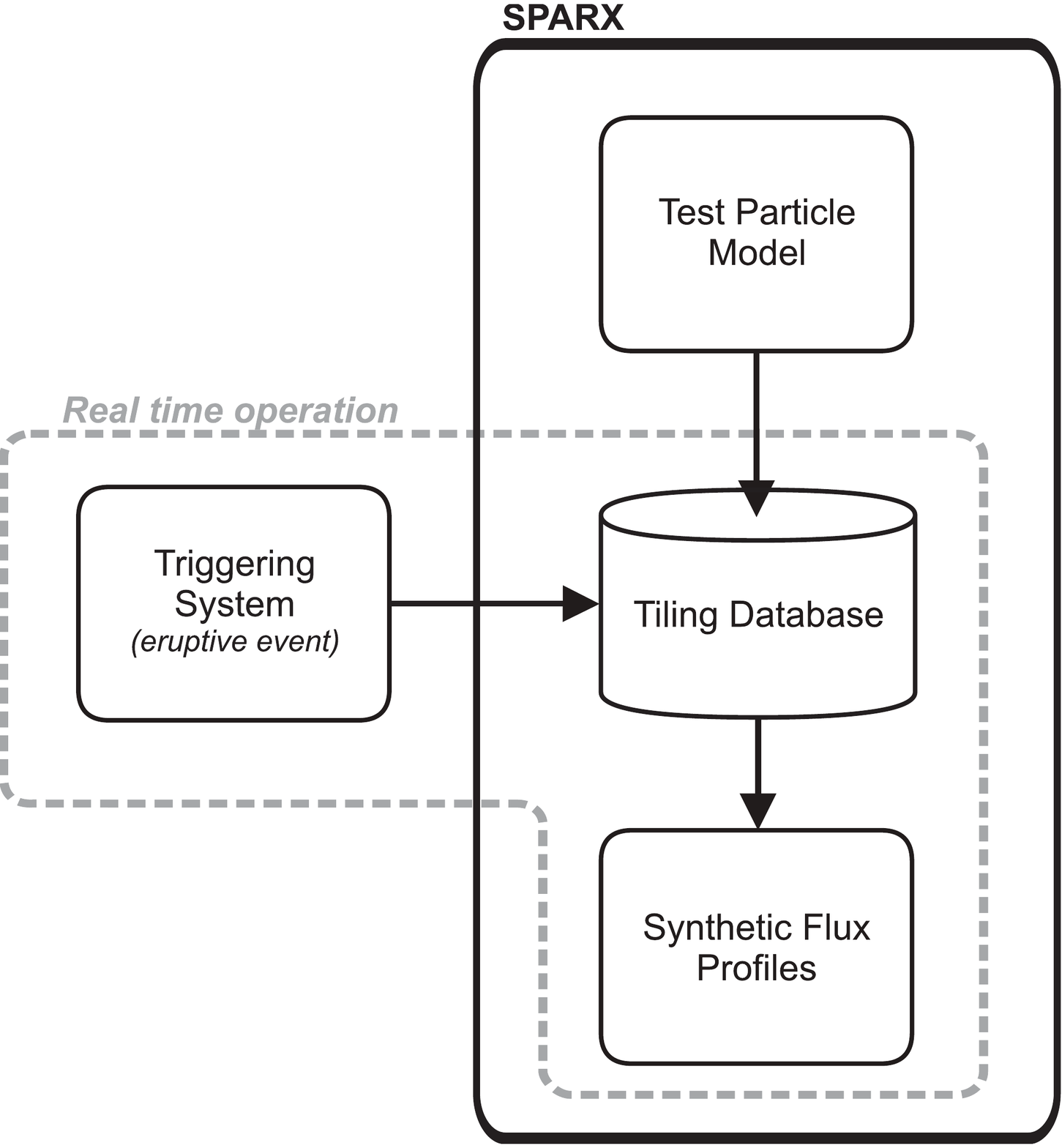}
\caption{Schematic illustrating the logical structure of the SPARX modelling system. 
The triggering system could be either an automated detection system for solar events (such as an X-ray flare as is the case within COMESEP) or manual input from a forecaster. The test particle model is a parallelised Fortran code. The tiling database methodology is described in Section~\ref{sec.tiling}. The visualisations and synthetic flux profiles presented in Figures~\ref{fig.proj} and \ref{fig.eastwest} are calculated by python post processing modules. 
}
\label{fig.flow}
\end{figure}

The SPARX forecasting system is extensible and can be modified to produce output from different particle injection configurations, injection spectra, output flux profile energy ranges, particle species, flux profiles at any point in the heliosphere, etc. The SPARX system may be run manually to produce a forecast of particle flux, here we present an example of its implementation within an operational forecasting system. 
SPARX is integrated within the SEP component of the operational COMESEP Alert System (www.comesep.eu/alert)
to produce a forecast of the time dependent flux of protons, for a given event, as would be detected in free space at 1~AU near the Earth-Moon system. Figure~\ref{fig.flow} shows a schematic representation, illustrating how the constituent components of the SPARX system are interrelated, as described in the following sections.

\subsection{Triggering system}

The SPARX system can be triggered manually by a forecaster using the location of a given event, here we describe its triggering within an automated operational system.
The COMESEP SEP module makes use of the results of an extensive statistical
analysis of SEP events and their associated solar eruptive events \cite{Die2015} as well as the output of SPARX.
Within the COMESEP Alert System, the SEP module is triggered by the real-time automated detection of a flare
by the Flaremail tool (sidc.oma.be/products/flaremail), which detects flares, and their peak flux, by
analysing the time profile of GOES Soft-X-Ray flux. The time of particle injection is taken to be the flare peak time, as identified by Flaremail.
The flare location on the solar disk is determined by the Solar DEMON system (solardemon.oma.be), by means of automated analysis of SDO/AIA 94~\mbox{\AA} images. When a flare is observed, SPARX is then triggered and the flare peak flux and location are passed as necessary inputs. Events beyond the limb cannot be used to produce a forecast in an automated operational system that relies on the flare site as the trigger, as implemented within the COMESEP alert system. However, these events could in principle be modelled by the SPARX system, to produce a forecast, if the SPARX code is run manually and assuming the source location and flare brightness can be estimated. It is possible that SPARX may be combined with probabilistic flare forecasts to produce an associated SEP forecast days in advance using the likelihood of observing each flare class for a given active region. It is possible to run the SPARX code to manually forecast the unnormalised flux profiles due to an event originating at any longitude at the Sun. 
In principle, SPARX could be triggered by any observed eruptive event, such as a CME or coronal dimming, using the observed location. The reason X-ray flares are currently used is to utilise the observed correlations between flare peak flux and proton peak flux to normalise the simulated flux profiles into physical flux units as described in Section~\ref{sec.profiles}. 
It is possible to use an observed CME as a trigger and a similar empirical relation between proton peak flux and CME parameters to normalise the simulated flux profiles. However, this approach has not been adopted due to the current large latency in obtaining coronagraph data and detection of a CME, compared to the short timescale between an event and SEP onset.

\subsection{Tiling database and optimisation} \label{sec.tiling}

Following the detection of a solar event, it is currently too time consuming to run the test particle model in real time considering the computational overheads; 
therefore, the approach adopted for operational running is to use a pre-generated database of model 
runs containing varying proton injection locations. 
The computational overhead of the test particle model and size of model database is greatly reduced by generating a database of results composed of small injection region tiles. 
This database is then queried in near-real-time following a solar event and the required tiles, dependent on the flare location, are combined to simulate a large extended region of particle injection.

In the current version of SPARX the database contains 30 model runs for each IMF polarity, each run describing a specific
6$^{\circ}$$\times$6$^{\circ}$ injection tile at a given central latitude $\psi_{c}$, central longitude $\phi_{c}=0^{\circ}$, and following the particles for 100 hours. 
The tiles have centers located at latitudes $\psi_{ci}=\pm \left( 3+6i \right)$ degrees, where $i=0,\ldots,14$.
Considering Eqs.(\ref{eqn_B})--(\ref{eqn_E}), the polarity of the magnetic and electric field contained within the model run tiles, used to produce a forecast for a particular event, is dependent on: 
\begin{equation}  
P\left(\theta\right) = A \, \mathrm{sgn} \left(\frac{\pi}{2} - \theta_{c} \right). \label{eqn_pol}
\end{equation}
The sign of $P\left(\theta\right)$ is determined by the event colatitude and magnetic phase of the solar cycle , where $A=1$ during an $A^{+}$ cycle, $A = -1$ during an $A^{-}$ cycle, sgn is the sign function, and $\theta_{c}$ is the event colatitude expressed in radians. In the current implementation, the polarities of all the tiles employed to produce a forecast for a particular event are the same, and are equal to that of the tile located closest to the event colatitude. 

To maintain a constant area density of particle injection within tiles centred at different latitudes, 
each injection tile models a total number of protons $N_{\psi_{c}}=N \cos(\psi_{c})$, where $N=100,000$.
The spectral index of the injection enegy spectrum is assumed to
be ${\gamma}$=1.1 and the mean free path $\lambda$=0.3 AU.
It is foreseen that, in the future, the database will be extended to allow for a wider range of variable parameters, and for
generating forecasts from an ensemble of runs.

By virtue of the rotational symmetry of the ${\bf B}$ and ${\bf E}$ fields, a model output database consisting of a single meridional strip of tiles may be used to construct a large particle injection region extended in longitude.
The output from the test particle model includes the three-dimensional coordinates of the particles, therefore, due to the rotational symmetry of the simple IMF model, the output obtained from the strip of injection tiles can be reproduced as if it were located at a different longitude, by rotating the relative longitude of the observer. In this way, it is possible to calculate the particle flux originating from the strip located at any longitude, using the model database containing a single meridional strip of injection tiles. The runs contained within the tiling database can then be used to construct an extended injection region of any width in longitude and latitude composed of 6$^{\circ}$$\times$6$^{\circ}$ tiles (see Figure~\ref{fig.injection}). This methodology provides the efficiencies required to calculate the proton flux profiles in an operational forecasting context.

\subsection{Construction of extended injection region}\label{sec.injection}

SPARX uses the information on the eruptive event location to
interrogate the database of model runs. Multiple injection tiles are combined and centered around the location
of the event, to simulate an
extended injection region that represents a CME-driven shock. 
The central latitude of the injection region is determined such that the subset of tiles extracted from the database, from the $\psi_{ci}$ latitude range described in Section~\ref{sec.tiling}, minimise the difference between the event latitude and central latitude of the injection region. 

\begin{figure}
\centering
\includegraphics[width=15pc]{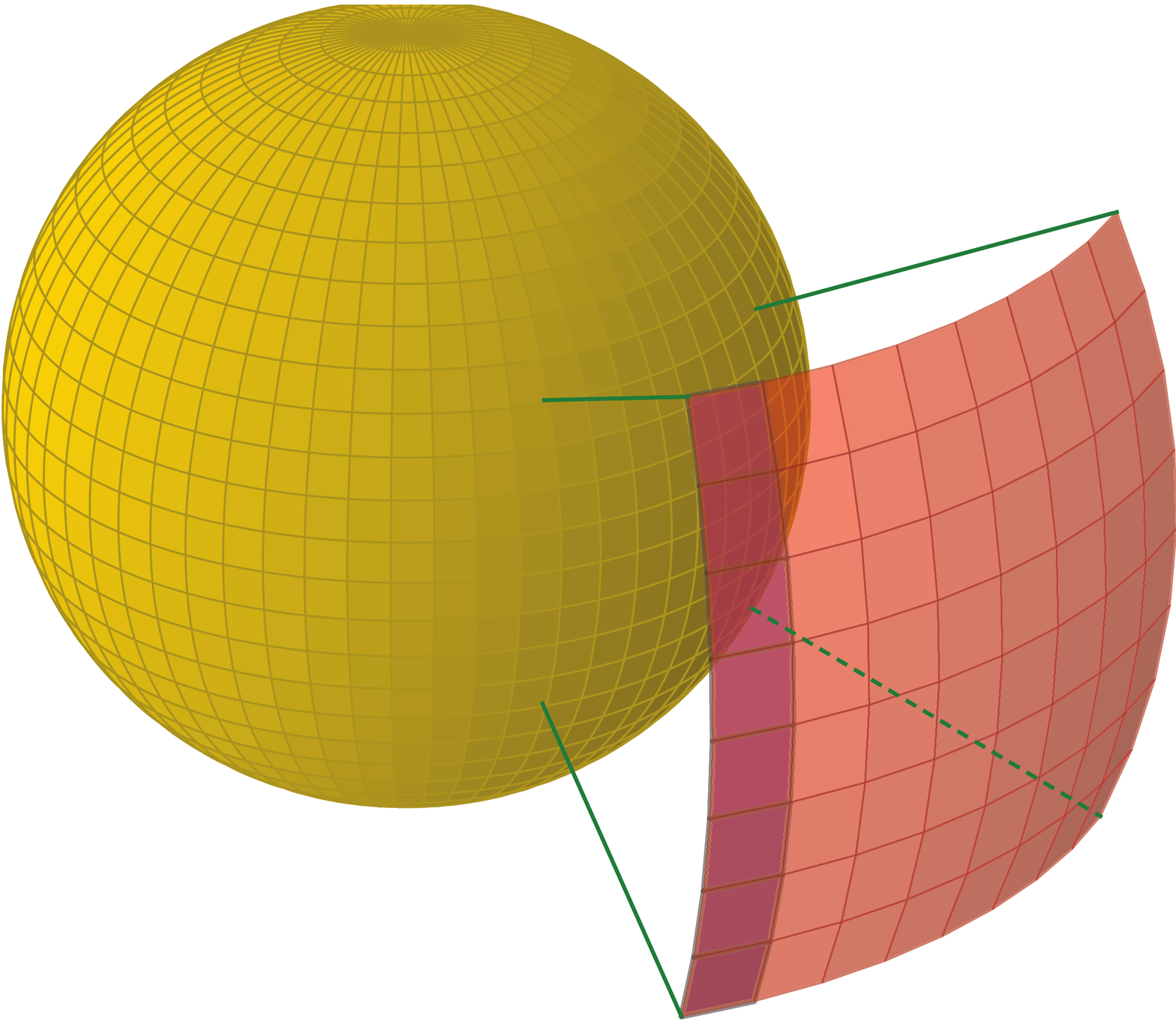}
\caption{Visualisation of how an extended particle injection region at 2~solar radii is constructed within the forecasting system. An extended injection region of 48$^{\circ}$$\times$48$^{\circ}$ in longitude and latitude is built up from 6$^{\circ}$$\times$6$^{\circ}$ model run tiles. Due to the rotational symmetry of the IMF model, a single meridional strip of run tiles contained within the SPARX test particle model database (darker shaded tiles) is used to build up an injection region extended in longitude, by a rotation of the relative longitude of the observer.}
\label{fig.injection}
\end{figure}

Figure~\ref{fig.injection} shows how the geometry of an extended particle injection region is produced. The injection region, located at 2~solar radii, is represented by a composite spherical grid of 8$\times$8 tiles. The darker shaded longitudinal strip of tiles represents the results contained within the model database.  
Given the current inability to reliably directly observe the shock front, and its extent, associated with a CME,
a fixed injection angular width of 48$^{\circ}$$\times$48$^{\circ}$ in longitude and latitude is assumed
for the operation of SPARX within the COMESEP Alert System; 
this is consistent with observed average CME properties \citep{Web2012}. 
In a non-automated run of SPARX it is possible for a forecaster to input the observed
CME shock extent.

In the current version of SPARX, the contribution of each tile is such that the area density of particle injection is uniform across the extended injection region, which corresponds to assuming that the efficiency of particle acceleration is constant along the shock front.

\subsection{Output: synthetic flux profiles}\label{sec.profiles}
The tiling database described in Section~\ref{sec.tiling} is comprised of the output from the test particle model and contains data on the time each particle crosses the sphere of radius 1~AU, its 3D coordinates, kinetic energy, and pitch angle. Therefore, the flux of particles through the surface at 1~AU can be computed. This is achieved by defining a surface area element through which the flux of particles can be integrated over specified time and energy ranges. The particle flux through this surface element can be calculated at any latitude and longitude, representing different observer spacecraft locations. To achieve a statistically significant number of counts, for operation within the COMESEP Alert System, the surface element is defined as 2$^{\circ}$$\times$2$^{\circ}$. 

The outputs of SPARX, within the COMESEP Alert System, are the near-Earth 
10-minute averaged integral proton flux profiles in the E$>$10 MeV and E$>$60 MeV ranges, which are comparable to those supplied by the GOES spacecraft. The time averaging period and energy ranges are easily customizable within the SPARX system and can be chosen to simulate the observations that would be made by any particular instrument.
To convert the flux of model particle counts into physical units, the peak proton fluxes are normalized depending on the flare peak X-ray flux. The normalization uses the correlations between solar event characteristics and
SEP peak fluxes described by \cite{Die2015}, where the average peak proton flux in five flare intensity bins is determined from a statistical survey of 90 SEP events, and associated flares, based on a subset of the SEPEM reference proton event list \citep[][Section 2.2, Table 9]{Die2015}.   
From the output of SPARX, the parameters of the predicted proton flux profiles such as the time to maximum flux, onset time, and duration are forecast at the Earth-Moon system.

\begin{figure*}[h]
\centering
\noindent\includegraphics[width=27pc]{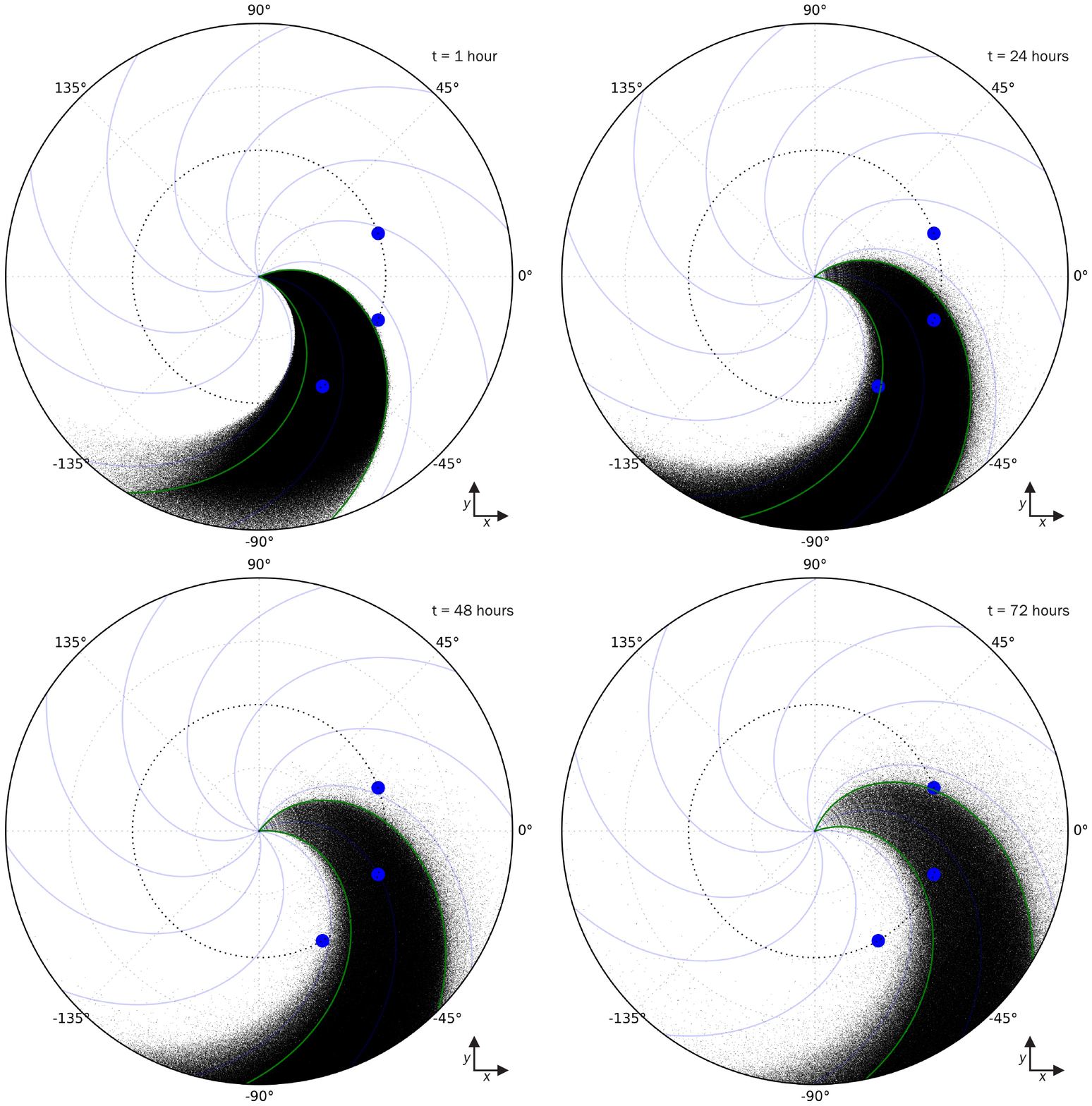}
\caption{Evolution of the Corotating Solar Energetic Particle Stream (CSEPS) produced by the SPARX test particle model (due to the particle injection region described in Section~\ref{sec.injection}) injected at N$20^{\circ}$ latitude. The evolution of the $10$-$400$~MeV proton particle stream is projected into the $x,y$ plane and shown at $t=$1, 24, 48, and 72 hours following the event injection (respectively: {\it top left, top right, bottom left, bottom right}). The equatorial IMF ({\it blue curves}), field lines bounding the injection region in the equatorial plane ({\it green curves}),  and orbit of radius 1~AU ({\it dotted circle}) are indicated. Three observer longitudes ($-60^{\circ}$, $-20^{\circ}$, and $20^{\circ}$) are shown in each panel at 1~AU ({\it blue circles}) representing the event viewed at relative locations to the Sun-observer line of ($\mathrm{W}60^{\circ}$, $\mathrm{W}20^{\circ}$, and $\mathrm{E}20^{\circ}$) respectively. The time evolution of the particle flux at these three locations is shown in the corresponding panels of Figure~\ref{fig.eastwest}. }
\label{fig.proj}
\end{figure*}

\begin{figure*}[h]
\centering
\noindent\includegraphics[width=42pc]{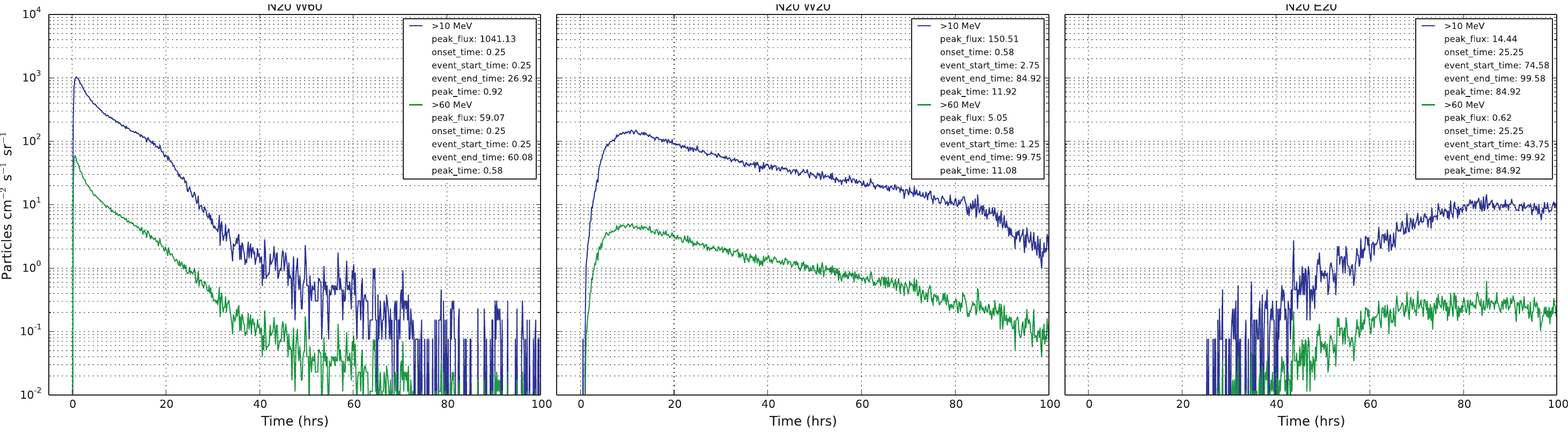}
\caption{Synthetic proton flux profiles produced by the SPARX forecasting system for an X10-class flare, accompanied by a CME, with the configuration shown in Figure~\ref{fig.proj}. The 10-minute averaged integral flux profiles for protons in the E$>$10 MeV ({\it blue curve}) and E$>$60 MeV ({\it green curve}) ranges are shown with event parameters listed ({\it insets}). {\it Left:} The flux profiles measured by an observer with the event originating at $\mathrm{W}60^{\circ}$ relative to the location of the Sun-observer line (corresponding to lower blue circle in each Figure~\ref{fig.proj} panel). {\it Middle:}  Flux profiles measured by the observer with the event at $\mathrm{W}20^{\circ}$ (middle blue circles in Figure~\ref{fig.proj}). {\it Right:}  Flux profiles measured by the observer with the event at $\mathrm{E}20^{\circ}$ (upper blue circles in Figure~\ref{fig.proj}).}
\label{fig.eastwest}
\end{figure*}

\section{Results} \label{sec.results}
To present the SPARX system, Figures~\ref{fig.proj} and \ref{fig.eastwest} demonstrate the output obtained for three different configurations of observer location relative to a simulated event, corresponding to an X$10$-class flare located at N$20^{\circ}$ latitude and $0^{\circ}$ longitude during an A$^+$ phase of the solar cycle. 

Figure~\ref{fig.proj} displays the output from the test particle model with the locations of  $\sim 6\times10^{6}$ protons that are initiated in this particular injection region, projected into the $x,y$ plane within 2~AU. 
The output is shown at 1, 24, 48, and 72 hours during the course of the event. It is apparent that the protons populate magnetic flux in a stream that corotates with the heliospheric magnetic field. We term this a Corotating Solar Energetic Particle Stream (CSEPS) which evolves over time, with the edges of the stream `softening' from a steep density gradient at the beginning of the event, to a more diffuse boundary of the particle stream as the event evolves. This dispersal of the particles in longitude is due to the effects of particle drifts described in \cite{Mar2013}.
Three observer longitudes at ($-60^{\circ}$, $-20^{\circ}$, and $20^{\circ}$) are indicated at 1~AU in Figure~\ref{fig.proj}, representing the event originating at ($\mathrm{W}60^{\circ}$, $\mathrm{W}20^{\circ}$, and $\mathrm{E}20^{\circ}$) relative to the Sun-observer line respectively.  Synthetic proton flux profiles are calculated at these three locations, to illustrate the effect on profile morphology, due to different observer longitudes within the CSEPS during an SEP event.

Figure \ref{fig.eastwest} shows the predicted proton flux profiles at 1 AU calculated by SPARX. The left panel shows the case where the event is Western with respect to the observer, originating at $\mathrm{W}60^{\circ}$ at the Sun with respect to the Sun-observer line; the middle panel shows a profile for the event at $\mathrm{W}20^{\circ}$ and the right panel for the event at $\mathrm{E}20^{\circ}$. The 10-minute averaged integral flux profiles for 
protons in the E$>$10 MeV and E$>$60 MeV ranges are shown.

The profiles shown in Figure~\ref{fig.eastwest} reproduce the well known East-West asymmetry 
in the morphology of SEP events \citep{Can1988}: for the case of the Eastern event ({\it right panel}) the rise time to peak is very slow, caused by the combined effect of the edges of the particle stream `softening' as the particle stream disperses in longitude due to drifts over time, and the time delay before the particle stream corotates over the location of the observer. The rise time decreases as the source region is positioned towards more Westerly locations.
For the case of W60 event ({\it left panel}), the rise time and overall duration of the event is much shorter, due to the direct magnetic connection with the injection region at the start of the event followed by the tendency for corotation to carry particles away from the observer as the event evolves. These trends are features that are observed in real SEP events \citep[cf.][Figure 2]{Can2006}.
Within the longitude ranges that are always directly magnetically connected to the initial injection region, the flux profiles are similar to the middle panel, due to the assumption of uniform shock injection efficiency currently used within SPARX. 

Drifts make the longitudinal range over which SEPs are detected within the IMF larger than the footprint of the injection region, and affect the flux profiles observed at 1 AU. Considering Figures~\ref{fig.proj} and \ref{fig.eastwest}, this is illustrated by the fact that the observer viewing the event at E20 relative to the Sun-observer line (Figure~\ref{fig.eastwest}: {\it right panel}) observes the onset at a time around 24~hrs after injection. Comparing this with the right hand panels in Figure~\ref{fig.proj}, if the particles were tied to the injection region field lines, we would not expect to observe the onset at this location until around 72~hrs. The longitudinal extent of the event is enhanced, due to drifts, with the result that the onset is observed much earlier at a location $\sim$30 deg from that expected under the field line tied paradigm described in Section~\ref{sect.intro}.
It should be emphasized that, in the absence of drift effects, the profile for the Eastern event (Figure~\ref{fig.eastwest}: right panel) would show a more impulsive profile with onset at $t \simeq 72$~hrs  when the injection region flux tube rotates past the observer; the slower rising profile is entirely due to the effects of drift. 

\section{Summary and conclusion} \label{sec.summary}

We present the first version of a new SEP propagation modelling system based on a full orbit
test particle approach, Solar Particle Radiation SWx (SPARX).
The model is three-dimensional and therefore able to describe particle transport across the magnetic field due to the
effect of drifts.

The model is currently operational within the COMESEP Alert System: triggered by the
automated detection of an X-ray flare, it constructs near-Earth SEP proton flux profiles.
The operational version is based on a database of model results, which are combined to construct an extended injection region representing a simplified CME shock, once the location of the flare is known.

In its initial form, the model includes instantaneous injection at the Sun from an extended shock-like region and uses a 
simplified configuration for the IMF polarity and constant solar wind speed. 

First results of SPARX show that the model predicts the range of magnetic field lines over which SEPs 
are detected is considerably larger than those originating at the footprint of the injection region, due to particle drifts.
The combined effect of drift and corotation is also observed to play a significant role in determining the longitudinal extent of SEP events and the observed flux profile. It is not, however, sufficient to explain the rapid spread of particles to very large longitudinal separations observed for example with STEREO \citep{Gom2015, Lar2013, Wie2013, Dre2014, Ric2014}. To explain the the latter observations the model may require the inclusion of more complex magnetic fields with the additional perpendicular transport that they produce via increased curvature and gradient drift \citep{Mar2013}, or field line meandering \citep{Lai2013}, but this is beyond the scope of the current version of the model. 

SPARX is able to reproduce qualitatively the well-known East-West variation of SEP profile morphology which depends on the relative longitude of the event with respect to the observer. Within the modelÕs assumptions, the primary cause of this East-West effect is due to corotation with the Sun of the magnetic flux, populated by the stream of SEPs, combined with drift effects. It appears that the effect of corotating SEP streams may be important in understanding the time dependent profiles of SEP events, although many models of SEP propagation are one dimensional and do not include the effect of corotation.
The importance of corotation was also noted by \cite{Dro2010}, who modelled STEREO SEP events using a 3D focussed transport approach. 

SPARX is a new modelling system that has transitioned the test particle approach from research to an operational forecasting context. Key features of the SPARX modelling system include: 
\begin{enumerate}
\item The role of drifts in determining the extent of regions populated by SEPs.
\item Deceleration associated with drift is fully taken into account including pitch-angle dependence \citep[see][]{Dal2015}.
\item It is possible to reproduce the East-West longitude dependence of SEP flux profile morphology without assuming a variation of injection efficiency along the shock; the primary cause of the East-West effect is corotating SEP streams (CSEPS). 
\end{enumerate}

We have presented SPARX, a physics based modelling system of SEP propagation that has been applied to produce rapid forecasts of the particle flux profiles at 1~AU, or elsewhere in the inner heliosphere. SPARX uses an innovative methodology to overcome the time constrained problem of physics based modelling of SEP flux profiles. This allows nowcasts and forecasts of SEP flux profiles to be produced within a timescale on the order of 2~mins from the SPARX code being triggered. The near real time functionality of SPARX requires a rather simplified configuration of the heliospheric fields, particle injection geometry, and assumed parameters such as the particle injection spectral index, scattering mean free path and constant solar wind speed. 
Future developments will include extending the number of variable parameters within
the database of model outputs (e.g. solar wind speed, mean free path, and injection spectrum) and making use of more accurate description of the IMF polarity, including the presence of the heliospheric current sheet. It is anticipated that SPARX will produce a forecast from an ensemble of runs with varying parameters, to assess the uncertainty of the forecast, adopting a similar approach to that used in meteorological forecasting.


\begin{acknowledgments}
This work has received funding from the European Union Seventh
Framework Programme (FP7/2007-2013) under grant agreement n.~263252
[COMESEP]. TL acknowledges support from the UK Science and Technology Facilities Council (STFC) (grant ST/J001341/1).
Data supporting Figures~\ref{fig.proj} and \ref{fig.eastwest} may be requested from the author M.S.~Marsh (mike.s.marsh@gmail.com).
\end{acknowledgments}

\bibliographystyle{agufull08}

\end{article}
\end{document}